# SIMULATED DYNAMICS AND ENDOHEDRAL RELESASE OF Ne FROM Ne@C$_{60}$ CLUSTERS

P. Tilton[1], B. Suchy[2], M.K. Balasubramanya[1] and M.W. Roth*,[2]

1) Department of Physical and Life Sciences, Texas A&M University-Corpus Christi, Texas 78412.
2) Department of Physics, University of Northern Iowa, Cedar Falls, Iowa 50614.

## 1. Abstract

Molecular Dynamics (MD) computer simulations are utilized to better understand the escape of neon from small ($N$=5) endohedral Ne@C$_{60}$ clusters. Multiple runs at various temperatures are used to increase the reliability of our statistics. The cluster holds together until somewhere between $T = 1150$K and $T = 1200$K, where it dissociates, showing no intermediate sign of melting or fullerene disintegration. When the temperature is increased to around $T = 4000$K, the encapsulated neon atoms begin to leave the cluster, with the fullerene molecules still remaining intact. At temperatures near $T = 4400$K, thermal disintegration of the fullerenes pre-empts the cluster dissociation. The neon atoms are then more quickly released and the fullerenes form a larger connected structure, with bonding taking place in atom pairs from different original fullerene molecules. Escape constants and half lives are calculated for the temperature range 4000K $\leq T \leq$ 5000K. The agreements and disagreements of results of this work with experiments suggest that classical MD simulations are useful in describing fullerene systems at low temperatures and near disintegration, but require more thought and modification before accurately modeling windowing at temperatures below $T = 3000$K.

*Corresponding author.  Fax; +1.319.273.7136 E-mail address: rothm@uni.edu (M.W. Roth)

## 2. Introduction

For two decades, fullerenes have remained at the forefront of scientific curiosity and research efforts. They are physically interesting molecules in their own right and can display a rich diversity of behavior when in the presence of other species. Because of their symmetry and durability it is of interest to place various atoms within a fullerene cage and look at physical changes that take place within the system. Many experimental[1-14] and computational[15-28] studies address the behavior of endohedral fullerene / noble gas systems as well as the dynamics of fullerene systems and how they may encapsulate certain atoms. Encapsulation of noble gases by fullerenes actually happens naturally, a fact which may yield significant information about certain aspects of the microscopic environment at the time of their formation. Specifically, helium isotopes may be trapped in fullerenes found in extraterrestrial objects or even on the ancient earth and, if the population is large enough, unique helium isotope ratios are preserved and can therefore be calculated.[29]

Another interesting aspect of encapsulation is that, for a cluster of Fulerene / endohedral noble gas molecules, the release of the encapsulated species can be studied as functions of temperature and time, possibly giving considerable insight into the age and local environment of a given system. Shimshi et al.[23] completed a mass spectroscopic study of the release of Ne@$C_{60}$. They found that it was possible for the fullerene to release a Ne atom without the molecular cage being destroyed, which is impossible if the Ne atom is simply pushed through the cage. Therefore they attribute the release of the encapsulated species to a window mechanism. They also find that, in the presence of impurities, the rate of release is increased by orders of magnitude. The half life for Ne escape at $T$ = 903 K is more than one month but at $T$ = 1173 K it is on the order of 10 hours. Here a modified window mechanism has been proposed, where the impurity (perhaps a radical) adds to the cage and weakens fullerene bonds and then leaves after the release, allowing reconstitution of the bond. The fullerene bonding topology and the integrity of the bonding has a profound effect on the system's dynamics, as the introduction of an ambient gas[1,4], the introduction of bonding atoms[31] or even trace amounts of impurities[30] can dramatically affect the release rates, almost always promoting more rapid release and at lower temperatures. It is even possible to experimentally hold a fullerene window open so as to facilitate endohedral capture and release[13].

It is clear that much remains to be understood about the endohedral release process. Since classical Molecular-Dynamics (MD) simulations have had some success in aiding our understanding of the dynamics exhibited by fullerenes and fullerene complexes, it seems reasonable to extend such simulations to study the dynamics of endohedral noble gas – fullerene clusters – in this case with the noble gas being Ne. Specifically, the purpose of this study is to understand the dynamics of Ne@$C_{60}$ clusters better, to unravel the endohedral release process, and to determine half – lives for release. Also, since many classical MD simulations of fullerene systems have been completed, this study will also facilitate a better understanding of their results and their range of applicability within the backdrop of agreement and conflict with experimental data.

### 3. Computational Approach

To simulate the Ne@$C_{60}$ complexes an $(N,V,T)$ Molecular Dynamics method is utilized. Initially, five endohedral molecules are placed near their equilibrium spacing in Fullerite. As simulated time runs forward, the equations of motion are integrated using a standard Verlet algorithm with a time step $\Delta t$=0.0005 ps, and various quantities of interest are calculated. At each of the temperatures $T$ = 1000K, 2000K, 3000K, 4000K and 5000K, 100 runs were completed, with initial velocities being randomized so that a more robust statistical picture of release can emerge from this study. In the temperature ranges 1000K $\leq T \leq$ 2000K and 4000K $\leq T \leq$ 5000K, the results of 5 different runs were averaged at temperatures spaced 50 K apart. The former range addresses cluster dissociation and the latter focuses on individual fullerene disintegration. Temperature control is achieved by velocity rescaling for the fullerene atoms and the noble gas population separately. Each run is taken out to $2 \times 10^6$ time steps (1 ns).

There are several types of interaction potentials used in the simulations. The Neon - Neon potential as well as the Neon-Carbon potential are of a Lennard-Jones form,

$$u_{LJ}(r_{ij}) = 4\varepsilon_{ij}\left[\left(\frac{\sigma_{ij}}{r_{ij}}\right)^{12} - \left(\frac{\sigma_{ij}}{r_{ij}}\right)^{6}\right], \quad (1)$$

where the potential parameters between various species are given in Table 1; mixed interaction parameters are obtained with the use of Lorentz-Bertholot combining rules involving C-C parameters for the same potential as in Equation 1. In addition there is a non-bonded carbon-carbon interaction [32] which is in a *modified* Lennard-Jones form

$$u_{LJ}(r_{ij}) = \varepsilon_{CC}\left[\left(\frac{\sigma_{CC}}{r_{ij}}\right)^{12} - 2\left(\frac{\sigma_{CC}}{r_{ij}}\right)^{6}\right]. \quad (2)$$

whose parameters are also shown in Table 1.

| Species | $\varepsilon_{ij}$(K) | $\sigma_{ij}$ (Å) |
|---|---|---|
| Ne-Ne | 36.68 | 2.79 |
| C-C* | 28.00* | 3.40* |
| C-C | 34.839 | 3.805 |

**Table 1.** Parameters for the non-bonded Lennard-Jones interaction potentials. The interactions with asterisks (*) in the middle row are not used explicitly in the simulations because they are for a standard LJ interaction, not the modified one actually used. Rather, they are used only in combining rule relationships to get Ne-C interactions of the form in Equation 1.

The carbon-carbon bonded interactions are modeled by Brenner's empirical extended bond-order potential [33]

$$\left.\begin{aligned}
V_R(r_{ij}) &= f(r_{ij})\frac{D_e}{S-1}\exp\left[-\beta\sqrt{2S}(r-R_e)\right] \\
V_A(r_{ij}) &= f(r_{ij})\frac{D_e S}{S-1}\exp\left[-\beta\sqrt{\frac{2}{S}}(r-R_e)\right] \\
f(r_{ij}) &= \begin{cases} 1, & r < R_1 \\ \frac{1}{2}\left[1+\cos\left(\frac{(r_{ij}-R_1)\pi}{(R_2-R_1)}\right)\right], & R_1 < r_{ij} < R_2 \\ 0, & r_{ij} > R_2 \end{cases}
\end{aligned}\right\}, \quad (3a)$$

which has parameters that are fit to various energetics of hydrocarbons, diamond and graphite. In Eqs. (3a), $V_R$ and $V_A$ are the repulsive and attractive terms, respectively, which are essentially modified Morse potentials. The screening function $f(r_{ij})$ restricts the interaction to nearest neighbors as defined by the values for $R_1$ and $R_2$. In addition, the

Brenner potential takes bonding topology into account with the empirical bond order function $\bar{B}_{ij}$ given by the relationships

$$\left.\begin{array}{l} B_{ij} = 1 + \left[ \displaystyle\sum_{k \neq i,j}^{N} G_C(\theta_{ijk}) f(r_{ik}) \right]^{-\delta} \\[1em] G_C(\theta) = a_0 \left[ 1 + \dfrac{c_0^2}{d_0^2} - \dfrac{c_0^2}{d_0^2 + (1+\cos\theta)^2} \right] \\[1em] \bar{B}_{ij} = \dfrac{B_{ij} + B_{ji}}{2} \end{array}\right\}. \quad (3b)$$

Here the three-body bond angle is defined as

$$\theta_{ijk} = \cos^{-1}\left( \frac{\vec{r}_{ji} \cdot \vec{r}_{jk}}{r_{ji} r_{jk}} \right), \quad (3c)$$

where $\vec{r}_{ab}$ is the displacement vector from carbon atom $a$ to carbon atom $b$. Variations of the Brenner potential have been used for many different types of carbon allotrope simulations, as the empirical bond order function controls clustering to some extent. For example, Yamaguchi et al.[22-26] do not include information from the conjugate bond-order term $B_{ji}$ because the potential would not adequately apply to small clusters. However, we are dealing initially with complete fullerenes, and therefore we include that term. The entire carbon-carbon interaction is a sum over all bonded and non-bonded interactions:

$$u_{CC} = \sum_{i=1}^{N} \sum_{j>i}^{N} \left\{ \left[ V_R(r_{ij}) - \bar{B}_{ij} V_A(r_{ij}) \right] + P_{ij} u_{LJ}(r_{ij}) \right\}. \quad (4)$$

Here $P_{ij}$ is a screening function [32] which we implement by creating bonded and non-bonded neighbor lists. All carbon-carbon bonded potential parameters are given in Table 2.

| Parameter | Value |
|---|---|
| $D_e$ | 73333.33 K |
| $\beta$ | 1.5Å$^{-1}$ |
| $S$ | 1.29 |
| $R_e$ | 1.315Å |
| $R_1$ | 1.750Å |
| $R_2$ | 2.000Å |
| $\delta$ | 0.80469 |
| $a_0$ | 0.011304 |
| $c_0$ | 19 |
| $d_0$ | 2.5 |

**Table 2.** Parameters for the bonded carbon-carbon Brenner interaction potential.

## 4. Results and Discussion

At low temperature ($T = 1000$K) the cluster holds together, as do the fullerene molecules comprising it. This is evidenced for the cluster by the sharp peak in the inter-fullerene pair distribution function $P_{inter}(r_{ij})$ shown in Figure 1, which is the calculated frequency of occurrence of a certain carbon atom pair separation between two carbon atoms in different fullerenes being between $r_{ij}$ and $r_{ij} + \Delta r_{ij}$, divided by normalizing factor $2\pi r_{ij}\Delta r_{ij}$.

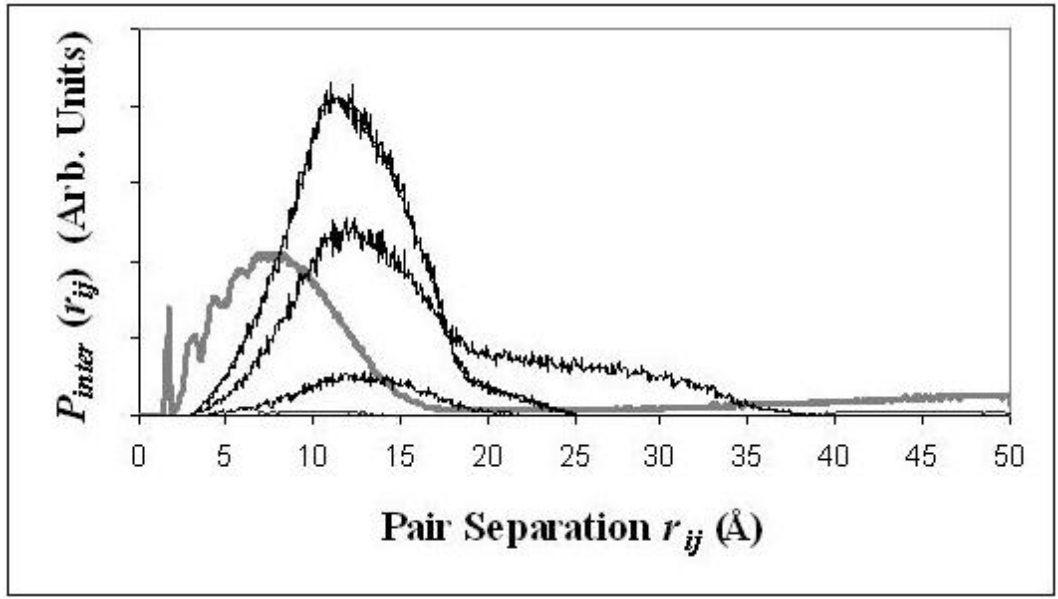

**Figure 1.** Inter-fullerene pair distribution function $P_{inter}(r_{ij})$. for temperatures $T = 1000$K and 1050K (the two nearly – coincident tallest bold curves), 1100K, 1150K, 1200K (the bold curve with the smallest peak) and 5000K (the light grey curve).

The low – temperature structure in the fullerenes themselves is quantified in a few different ways. Figure 2 shows a sharp low – temperature peak in $P(r)$, the fullerene radial probability distribution, which is the calculated probability of a carbon atom being a certain distance from the center of mass of the fullerene molecule containing it. Moreover, the low – temperature fullerene pair distribution function $P_f(r_{ij})$ shown in

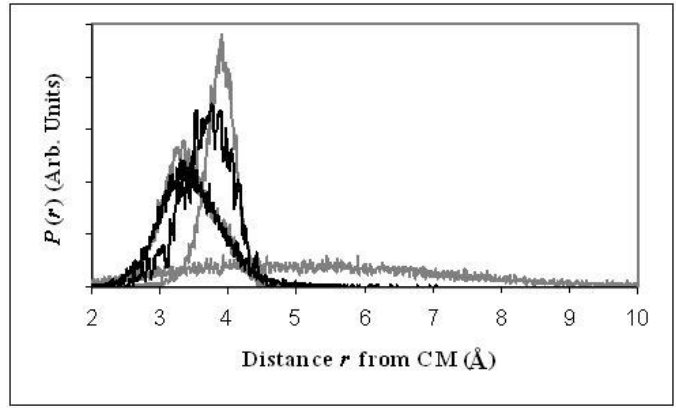

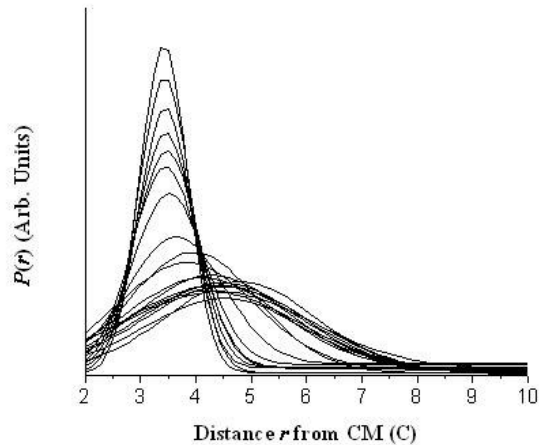

**Figure 2.** *P(r)*, the fullerene radial probability distribution for $T$ = 1000K, 2000K, 3000K, 4000K and 5000K (upper plot with alternating grey and black curves), and best Gaussian fits for 4050K $\leq T \leq$ 4950K (lower plot). The Gaussian fits are chosen for visualization ease because of the number of curves present, and the fastest change in their character take place between $T$ = 4300K and $T$ = 4500K.

Figure 3 exhibits a strong solid – like signature. $P_f(r_{ij})$ is calculated exactly as $P_{inter}(r_{ij})$ is, with the exception that both carbon atoms in the pairs are necessarily in the same fullerene molecule.

Inspection of $P_{inter}(r_{ij})$ in Figure 1 gives an indication of the first significant change in the system with increasing temperature. As the temperature increases from $T$ = 1000K, the peak near $r_{ij}$ = 12 Å dramatically drops in amplitude due to thermal fluctuations. By $T$ = 1200 K the distribution is essentially flat, because all the atom pair separations are greater than the highest value of the abscissa of the plot. Therefore the data seems to suggest that the cluster dissociates in the temperature range 1150K ≤ $T$ ≤ 1200K, which is in reasonable agreement with other calculations [34,35] of the melting of C60 structures under pressure, and the well known fact that, at low pressure, fullerites sublime and do not melt. The plots of $P_f(r_{ij})$ in Figure 3 illustrate that the dissociation of the cluster has no noticeable general effect on the structure of the fullerene molecules themselves. However the plateau between $r_{ij}$ = 6 Å and 7.5 Å present at low temperature dies off more quickly at higher temperatures, suggesting that the walls of the molecules themselves become slightly thinner with increasing temperature. The plots of $P(r)$ in Figure 2 confirm the insensitivity of the fullerene molecules to the cluster dissociation and also show that in the temperature range 1050K ≤ $T$ ≤ 1950K the peak in the distributions actually shifts towards lower distances, which suggests that as temperature increases from $T$ = 1000K the molecules also contract slightly.

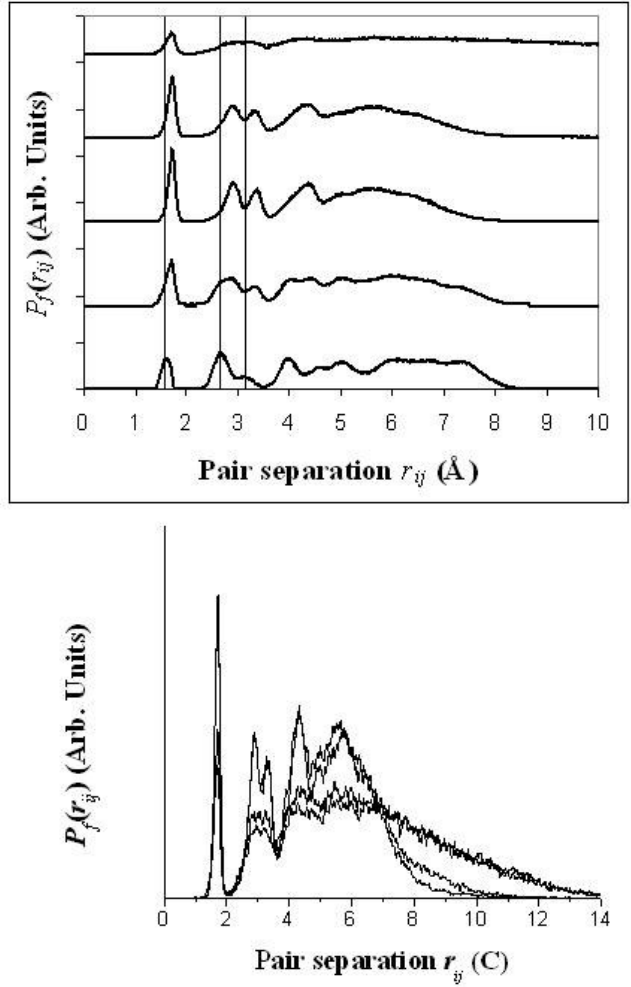

**Figure 3.** fullerene pair distribution function $P_f(r_{ij})$ for $T$ = 1000K, 2000K, 3000K, 4000K and 5000K (upper plot) and for $T$ = 4050K, 4300K, 4500K and 4950K (lower plot). In the upper plot, vertical lines indicate the location of the low – temperature peaks for first, second, and third neighbors. In the lower plot, the solid – like curves with sharp peaks are for the two lower temperatures and the lower curves showing much less structure are for the higher two temperatures.

As the temperature is raised beyond $T$ = 2000K, the system shows expected signs of thermal fluctuation, with very little change between $T$ = 3000K and $T$ = 4000K, and fullerene molecules still remaining intact. This is evidenced in part by the profound solid – like character for the system still present in the distributions at $T$ = 4000K and 4500K shown in Figures 2 and 3. In addition, the thermal average of the system's total configurational energy $<U_c>$ $(T)$ shown in Figure 4 does not exhibit a sharp increase, showing the absence of a proliferation in carbon – carbon bond breaking which would be present for disintegration.

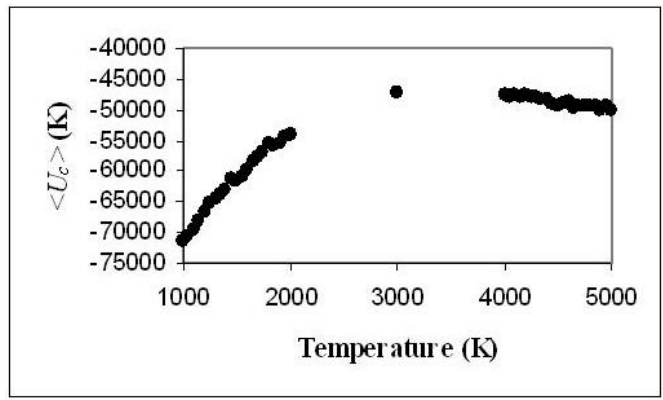

**Figure 4.** Thermal average of the system's total configurational energy, $<U_c>$ $(T)$.

Moreover, the average maximum and minimum moments of inertia ($I_{max}$ and $I_{min}$, respectively, shown in Figure 5) are calculated by obtaining the maximum and minimum eigenvalues of the inertia tensor for each molecule and averaging them over all the runs at a given temperature.

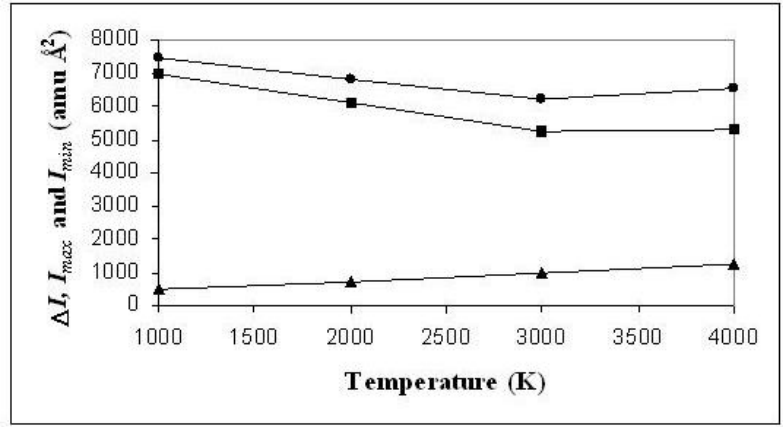

**Figure 5.** Maximum (circles) and minimum (squares) moments of inertia $I_{max}$ and $I_{min}$, respectively, for the fullerene molecules, and the difference $\Delta I$ between the two extreme inertia values (triangles).

The two extreme moments of inertia become slightly more extreme as temperature increases, which is expected. The difference $\Delta I$ between the two extreme moments of inertia (a quantitative measure of the degree of oblateness of the fullerenes) increases with temperature but does not show any sharp changes at or below $T = 4000K$, confirming that the fullerene molecules remain intact up to this temperature.

At $T \approx 4000K$, the first endohedral neon atoms are released from the cluster. As the temperature increases, atom release begins sooner and happens more rapidly. This is evidenced in part by Figure 6, showing $T_1$, the time of release for the first atom in the cluster, as a function of temperature. In addition, the number of Ne atoms encapsulated within the cluster as functions of time for each temperature is fit to curves of the form

$$N(t,T) = N_0 e^{-k(T)t}, \qquad (5)$$

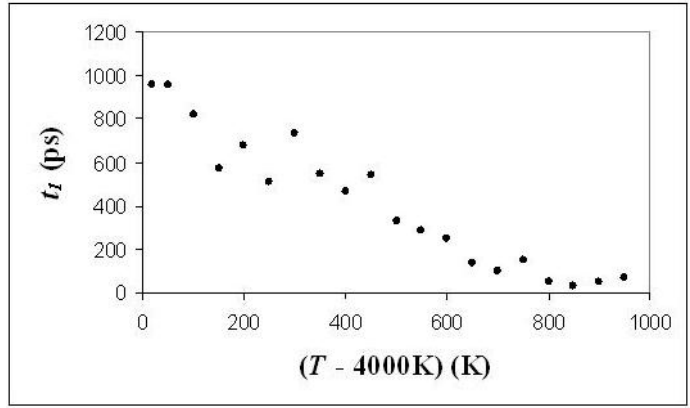

**Figure 6.** Time $t_1$ of first atom release as a function of temperature. The abscissa is offset by 4000K in order be consistent with that of Figure 7.

and the rate constants *k(T)* as functions of temperature are shown in Figure 7.

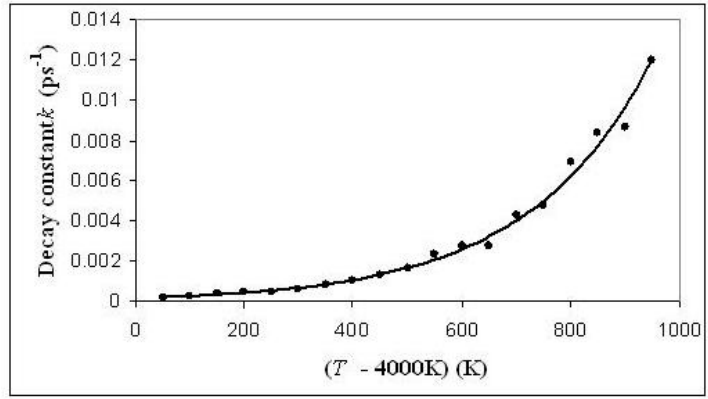

**Figure 7.** Rate constants *k(T)* as functions of temperature. The solid line is the best exponential fit to the data (for visual aid) and the abscissa is offset by 4000K in order to facilitate curve fitting.

Selected release half lives as calculated from the rate constants are shown in Table 3.

| Temperature | $t_{1/2}$ | Source |
|---|---|---|
| 4050 K | 2.97 ns | This work |
| 4500 K | 0.41 ns | This work |
| 4950 K | .057 ns | This work |
| 903 K | 6.1 yr | [30] |
| 1173 K | 57.8 h | [30] |
| 293 K* | 40.3 h* | [13]* |

**Table 3.** Half lives $t_{1/2}$ for endohedral escape for Ne and, in the bottom row (*), for He.

As the temperature increases past $T = 4000K$, $P(r)$ in Figure 2 and $P_f(r_{ij})$ in Figure 3 show that the fullerenes themselves are losing structural order. In fact, the curves for the finer temperature scale ($4050K \leq T \leq 4950K$) in Figure 2 and the bottom plot in Figure 3 both show that the fullerenes lose a considerable amount of structural order vis-à-vis the radial atomic distribution distance as well as the pair distribution function, respectively, near $T = 4400K$. Since this is when the moments of inertia begin to grow very large (well off the scale of the plot) but the configurational energy does not, the loss of structural order here is interpreted to be from disintegration and recombination. The temperature at which the fullerenes themselves show this profound loss of structural order is in excellent agreement with disintegration temperatures obtained by other MD calculations involving $C_{60}$.[19] In contrast, between $T = 4000K$ and $4300K$ it seems that windowing is taking place, *with permanent disorder imparted to the fullerene*. Hence, our simulations strongly support the idea that impurities catalyze some windowing mechanism in the fullerene. The presence of recombination is also confirmed by the $P_{inter}(r_{ij})$ curve at $T = 5000K$ in Figure 1, where there is a sharp peak near 1.5 Å indicating that bonding *between atoms from different fullerenes* has taken place. The series of broader peaks at higher separations shows that atoms from different fullerenes have merged to create other bonded structures. Inspection of the structures shown at the end of the simulations indeed confirms that, if the fullerenes themselves exhibit disorder before the cluster dissociates, then recombination can occur.

      The Ne@C60 dissociation temperature, as well as the temperature of thermal disintegration of our fullerene molecules, are in good agreement with previous calculations. [19, 34, 35] Although we are unable to locate any prior simulation studies of

endohedral noble gas atom release, the results obtained in this study differ *considerably* from that obtained experimentally [30] and some discussion follows. Our simulations are conducted with no species of any kind surrounding the fullerenes, and the only way bonds can be broken is with thermal agitation – i.e. the character of the C-C bonds are not being altered. This is in contrast to other simulations where higher energy noble gas atoms collide with the fullerene structure, causing a window which then closes.[15,16] In real experiments there are other atoms / molecules surrounding the system, and bond breaking / forming has been shown to be extremely sensitive to any species that may attempt to bond with the fullerene, and the effect is almost always to accelerate release. It may be that, in the real molecule, interactions of the outer electrons of noble gas atoms encapsulated within the fullerenes interact dynamically with the fullerene bonds. In fact, when a window is held open in a fullerene, $^3$He escape half lives at $T = 333$K are already as low as 1.4 hours. In our study, then, a picture emerges where the simulation results seem generally reasonable at low temperatures where non – bonded Lennard – Jones interactions dominate phase transitions, and near thermal disintegration of fullerenes. Hence, in the temperature range 4000K $\leq T \leq$ 5000K, it seems reasonable to assume that the windowing suspected in causing the Ne release seen in experiment [30] would not have nearly the effect that thermal windowing seen in simulations does, and our escape calculations and any comparisons across the noble gases may be generally applicable as well. However, in order to accurately deal with thermal windowing and model escape at temperatures well below those for thermal disintegration, two challenges arise immediately. To begin with, a temperature – dependent description of the bond integrity as affected by any exohedral species will be required. That is, the simulations should somehow take into account the effect of impurities on any windowing mechanisms. Moreover, the escape process will have to be artificially accelerated in such a way as to preserve the integrity of the physics of the model, because in order for the endohedral population to become half of its initial value, the simulation will have to run for one half life which, even taking an underestimate of $t_{1/2}$ = 10 hours, would require 7.2 x $10^{19}$ steps using a time step of 0.0005 ps, translating to a real time commitment (in the case of this study) of more than the currently accepted value for the age of the universe.

## 5. Acknowledgements

M.W.R acknowledges useful discussions with J. Che and R.J. Cross, as well as a UNI student fellowship for B.S. during the summer of 2004. M.K.B. acknowledges support for Molecular Dynamics simulations by the National Science Foundation's MRI program, under the Division of Computer and Network Systems' grant numbered 0321218.